\documentclass{optica-article}

\journal{opticajournal} 

\articletype{Research Article}

\usepackage{lineno}
\usepackage{graphicx}
\usepackage{makecell}
\usepackage{amsmath}
\usepackage{amsfonts}
\usepackage{multirow}
\usepackage{array}
\usepackage{enumerate}
\usepackage{stfloats}
\usepackage{cite}
\usepackage{booktabs}
\usepackage{bm}

\begin{document}

\title{Layered soft-output MLSE architecture for PAM4 short-reach optical interconnects}

\author{Mengqi Guo,\authormark{1} Ji Zhou,\authormark{2,*} Zhengxiang Cheng,\authormark{3} Changyuan Yu,\authormark{4} Xiangjun Xin,\authormark{2} and Liangchuan Li\authormark{2}}

\address{\authormark{1}School of Information Science and Technology, Shijiazhuang Tiedao University, Shijiazhuang 050043, China\\
\authormark{2}Aerospace and Informatics Domain, Beijing Institute of Technology, Zhuhai 519088, China\\
\authormark{3}Department of Electronic Engineering, College of Information Science and Technology, Jinan University, Guangzhou 510632, China\\
\authormark{4}Department of Electrical and Electronic Engineering, The Hong Kong Polytechnic University, Hong Kong SAR, China
}

\email{\authormark{*}zhouji@bitzh.edu.cn} 

\begin{abstract*}
The rapid growth of data traffic driven by artificial intelligence demands high-speed data-center optical interconnects, where soft-output maximum-likelihood sequence estimation (SMLSE) provides reliability information for soft-decision forward error correction. However, conventional SMLSE suffers from a latency bottleneck because sequential state-metric recursions cause latency to increase linearly with the number of symbols. Here, we propose a layered SMLSE architecture that reduces latency from linear to logarithmic order. Downward branch-metric accumulation combines adjacent interval metrics in parallel, while upward state-metric propagation reconstructs forward and backward state metrics at successive interval midpoints. Experiments with 112-Gbit/s PAM4 transmission over 2-km standard single-mode fiber demonstrate identical soft-output performance for layered and conventional SMLSE, while the total latency decreases from 26 to 10 processing cycles. The proposed layered SMLSE provides a low-latency architecture while preserving soft-output performance for high-speed optical interconnects.

\end{abstract*}

\section{Introduction} \label{section1}
Large-scale distributed artificial intelligence training and inference generate massive inter-server traffic, driving stringent requirements for high-speed and low-latency optical interconnects in data centers \cite{saber2025physical}. For short-reach optical interconnects, intensity modulation and direct detection (IM/DD) remains attractive owing to its cost and implementation advantages \cite{Oettinghaus2026DSP400G}. However, the growing mismatch between increasing symbol rates and limited transceiver bandwidth results in severe intersymbol interference (ISI), motivating the use of receiver-side equalization and sequence detection techniques, including feed-forward equalizers (FFEs), decision-feedback equalizers (DFEs), and maximum-likelihood sequence estimation (MLSE) \cite{Che2024Modulation,Nagarajan2024LowPowerDSP,zhou1,zhou2,Wu2024GradientDescent,Rha2025EnhancedDFEMLSE}. By performing maximum-likelihood detection over candidate symbol sequences, MLSE achieves optimum sequence detection performance in ISI channels \cite{Forney1972MLSE}. Meanwhile, soft-decision forward error correction (SD-FEC) is becoming increasingly important in next-generation high-speed IM/DD optical interconnects, with soft-decision inner-FEC decoding being considered in IEEE P802.3dj \cite{IEEE8023dj}. Since conventional Viterbi-based MLSE provides only hard decisions, soft-output MLSE (SMLSE) is required to generate the soft information needed for SD-FEC decoding \cite{Djordjevic2009NextGenerationFEC}.

Existing studies on SMLSE, typically implemented using Bahl-Cocke-Jelinek-Raviv (BCJR) or its Max-Log-MAP approximation \cite{Bahl1974BCJR,Robertson1997OptimalMAP}, mainly focus on improving soft-output performance and reducing computational complexity. Least-square-assisted BCJR improves soft-output detection by enhancing branch-metric estimation under severe ISI \cite{Liu2025LSBCJR}, and SMLSE has been combined with SD-FEC to improve decoding performance \cite{Moon2019SoftOutputMLSE}. To reduce computational complexity, M-algorithm-based state reduction has been employed in BCJR for bandwidth-limited PAM4 interconnects \cite{Zou2023THPMlogMAP}. FFE- and DFE-assisted SMLSE schemes use equalizer outputs to guide trellis-state reduction \cite{jing,Zhang2025LowComplexityMLSE}. A state-integrated look-up table can reduce redundant metric calculations in BCJR \cite{Liu2025SiLUTBCJR}. Traceback-assisted SMLSE eliminates the backward recursion using traceback information \cite{Zhao2025TracebackAssistedMLSE}. However, reducing states or arithmetic operations does not necessarily reduce processing latency. The inherent symbol-by-symbol recursion in SMLSE results in a long sequential processing dependency, making processing latency increasingly critical as symbol rates continue to increase. Our previous work reduced hard-output detection latency from linear to logarithmic order for MLSE \cite{Guo2026LayeredMLSE}, but did not consider the latency bottlenecks associated with bidirectional state-metric recursion and soft-information generation in SMLSE.

\begin{figure}[!t]
\centering
\includegraphics[width=11.2cm]{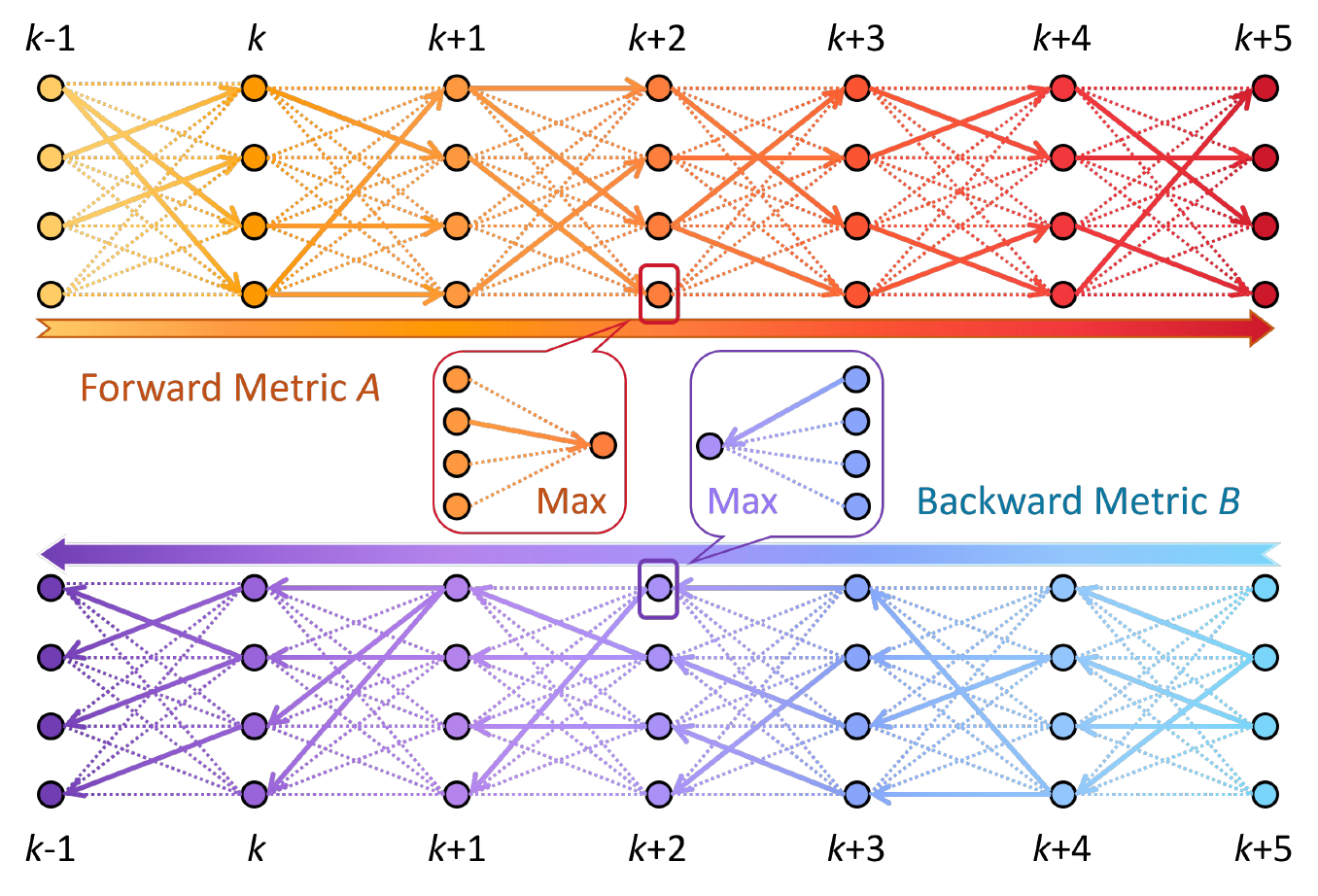}
\caption{Forward and backward state-metric recursions in conventional SMLSE for PAM4 signals. Solid lines indicate the selection of the maximum state metrics.}
\label{fig:conventional_smlse}
\end{figure}

To enable low-latency SMLSE, we propose a layered SMLSE architecture for PAM4 short-reach optical interconnects. The architecture reorganizes the bidirectional state-metric recursion into downward branch-metric accumulation and upward state-metric propagation. Adjacent interval metrics are first combined in parallel, after which the stored metrics are used to reconstruct forward and backward state metrics at successive interval midpoints. Layered SMLSE reduces the processing latency from linear to logarithmic order while preserving the soft outputs of conventional SMLSE. Experimental results for 112-Gbit/s PAM4 transmission over 2 km of standard single-mode fiber (SSMF) show that the proposed layered SMLSE and conventional SMLSE have identical bit-error rate (BER), normalized generalized mutual information (NGMI), and bit-level log-likelihood ratios (LLRs). Assuming that each recursion step or parallel processing layer takes one processing cycle, the estimated latency decreases from 26 to 10 processing cycles. The proposed layered SMLSE offers a promising approach to low-latency soft-output detection for SD-FEC in next-generation data-center optical interconnects.


\begin{figure}[!t]
\centering
\includegraphics[width=10.9cm]{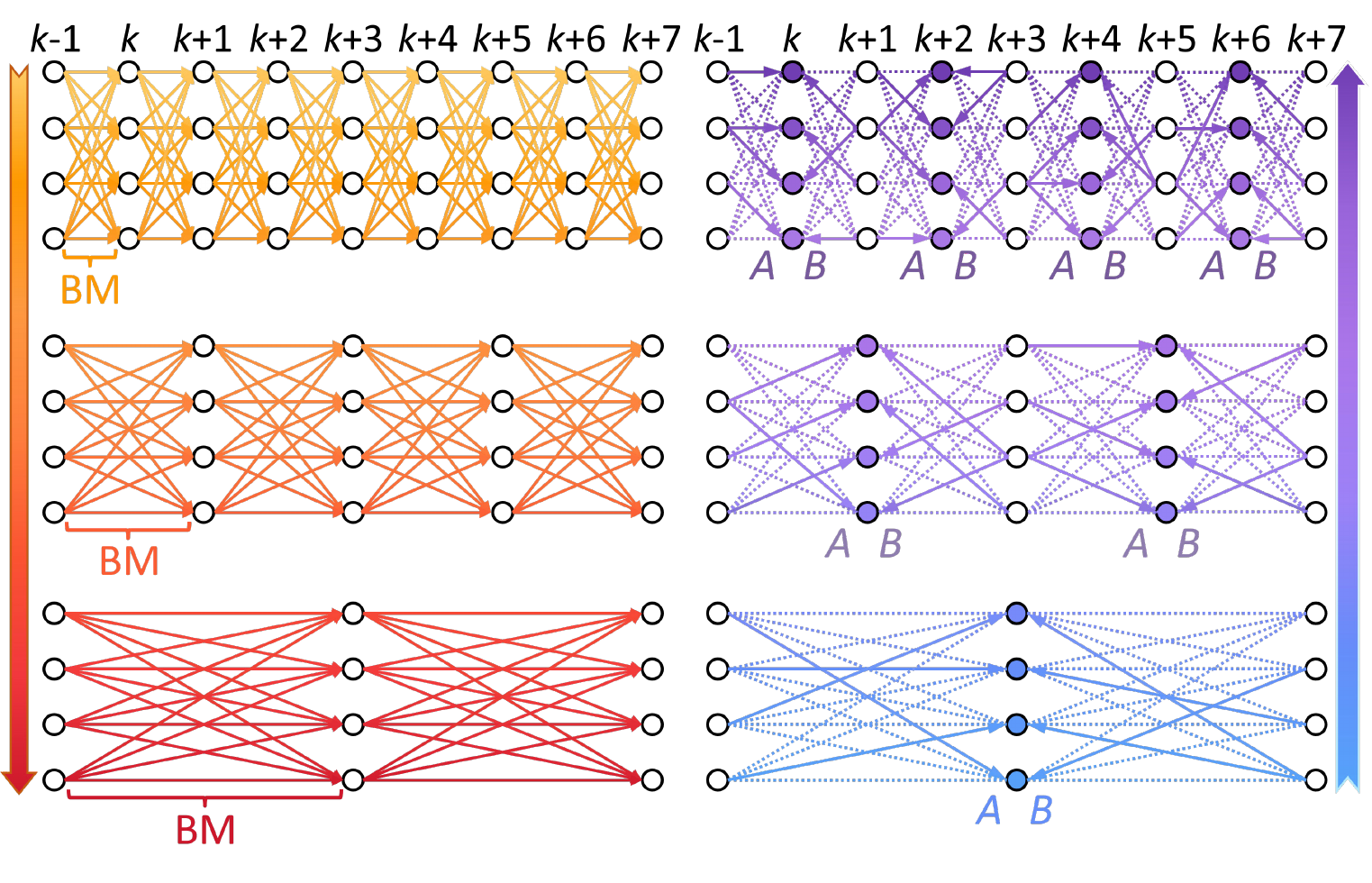}
\caption{Downward branch-metric accumulation and upward state-metric propagation in the proposed layered SMLSE for PAM4 signals. BM: branch metric.}
\label{fig:proposed_smlse}
\end{figure}

\section{Principle}
\label{sec:principle}

This section briefly introduces the principles of conventional SMLSE and then presents the proposed layered SMLSE architecture. Conventional SMLSE employs forward and backward state-metric recursions, whereas the proposed layered SMLSE uses downward branch-metric accumulation followed by upward state-metric propagation. This layered processing reduces the processing latency from linear to logarithmic order.

\subsection{Conventional soft-output MLSE}
\label{sec:conventional_smlse}
The conventional SMLSE employs the Max-Log-MAP algorithm, which reformulates BCJR in the logarithmic domain and uses the max-log approximation to reduce computational complexity~\cite{Zhang2025LowComplexityMLSE}. Its processing flow and state transitions are illustrated in Fig.~\ref{fig:conventional_smlse}. For a processing block containing $N$ symbols, the forward state metric $A_k(v)$ and the backward state metric $B_{k-1}(u)$ are computed recursively under the max-log approximation as
\begin{equation}
  \begin{aligned}
    A_k(v)
    =\max_{u,v\in\mathcal X}
      \left\{A_{k-1}(u)+\Gamma_k(u,v)\right\},\\
    B_{k-1}(u)
    =\max_{u,v\in\mathcal X}
      \left\{\Gamma_k(u,v)+B_k(v)\right\},
  \end{aligned}
  \label{eq:conventional_recursions}
\end{equation}
where $k=1,\ldots,N$ and $\mathcal X=\{-3,-1,+1,+3\}$ is the PAM4 symbol set. The variables $u$ and $v$ denote the trellis states at time indices $k-1$ and $k$, respectively. $\Gamma_k(u,v)$ is the branch metric associated with the transition from state $u$ to $v$. For a two-tap post-filter with coefficients $h_0$ and $h_1$, the branch metric is expressed as
\begin{equation}
  \Gamma_k(u,v)
  =-\frac{\left|y_k-\left(h_0v+h_1u\right)\right|^2}{2\sigma^2},
  \label{eq:branch_metric}
\end{equation}
where $y_k$ is the received symbol at time $k$, $h_0$ and $h_1$ are the post-filter coefficients, and $\sigma^2$ is the noise variance. In general, $h_0$ is normalized to unity, and $h_1$ is denoted by $\alpha$ \cite{hoon}. Therefore, the PAM4 trellis contains four states and sixteen possible transitions per symbol interval.

The forward and backward state metrics are combined to estimate the posterior probability of each transmitted symbol. The logarithmic symbol posterior under the max-log approximation is expressed as
\begin{equation}
  \ln P(x_k=v\mid\mathbf y) = A_k(v)+B_k(v),
  \qquad v\in\mathcal X,
  \label{eq:state_soft_metric}
\end{equation}
where $x_k$ is the transmitted symbol at time $k$ and $\mathbf y=[y_1,\ldots,y_N]$ is the vector of received symbols. The forward metric accounts for the received symbols up to time $k$, whereas the backward metric accounts for those after time $k$. Their combination incorporates observations from both directions. For Gray-coded PAM4, the levels $\{-3,-1,+1,+3\}$ are assigned the labels $\{00,01,11,10\}$, respectively. Defining the bit LLR as the logarithm of the probability ratio of bit~1 to bit~0, the LLRs of the most significant bit (MSB) and the least significant bit (LSB) are given by
\begin{equation}
  \begin{aligned}
    L_{k,\mathrm{MSB}}
    &=\ln
    \frac{P(x_k=+1\mid\mathbf y)+P(x_k=+3\mid\mathbf y)}
         {P(x_k=-3\mid\mathbf y)+P(x_k=-1\mid\mathbf y)},\\
    L_{k,\mathrm{LSB}}
    &=\ln
    \frac{P(x_k=-1\mid\mathbf y)+P(x_k=+1\mid\mathbf y)}
         {P(x_k=-3\mid\mathbf y)+P(x_k=+3\mid\mathbf y)}.
  \end{aligned}
  \label{eq:pam4_bit_llrs}
\end{equation}
In the Max-Log-MAP implementation, the logarithm of each probability sum is approximated by the largest logarithmic symbol probability. The sign of the LLR indicates the preferred bit value, while its magnitude reflects the reliability of the corresponding decision.

\subsection{Proposed layered soft-output MLSE}
\label{sec:layered_smlse}

\begin{figure}[!t]
\centering
\includegraphics[width=11cm]{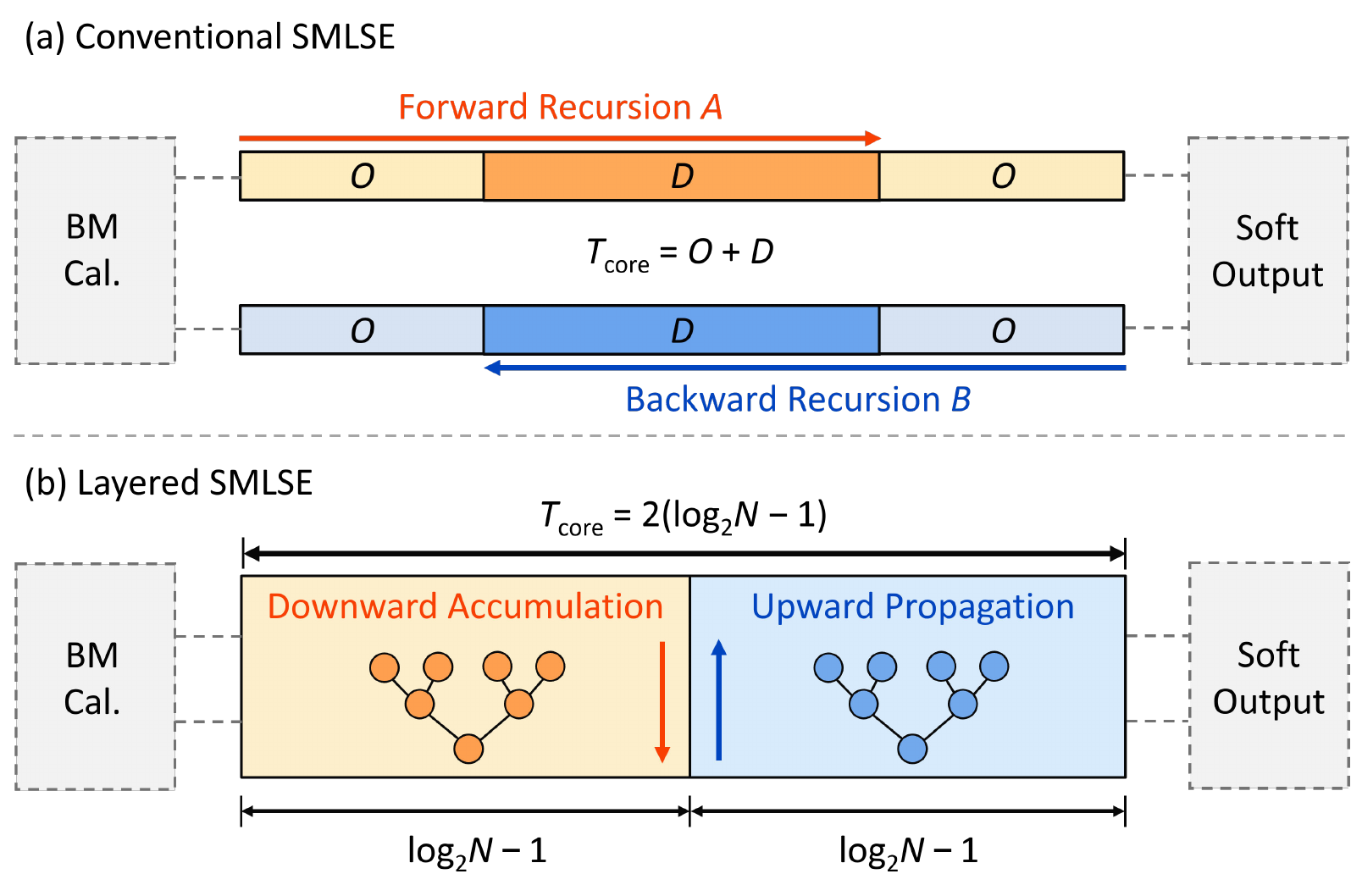}
\caption{Processing schedules and core latency of (a) conventional SMLSE and (b) the proposed layered SMLSE. BM Cal.: branch-metric calculation.}
\label{fig:processing}
\end{figure}

The proposed architecture reorganizes the sequential metric computations into a layered structure, as shown in Fig.~\ref{fig:proposed_smlse}. The layered structure comprises a downward branch-metric accumulation stage and an upward state-metric propagation stage. In the downward stage, adjacent interval metrics are combined, and their accumulated metrics are stored. In the upward stage, the stored interval metrics are used for state-metric propagation. The forward metric $A$ is propagated from the left boundary to the midpoint of each interval. The backward metric $B$ is propagated from the right boundary to the same midpoint of each interval.

Consider an interval $[a,b]$ within a processing block containing $N$ symbols, where $0\le a<b\le N$. The optimum accumulated metric between state $u$ at time $a$ and state $v$ at time $b$ is defined as
\begin{equation}
  G_{a:b}(u,v)=
    \max_{\substack{s_{a+1},\ldots,s_{b-1}\in\mathcal X\\
                     s_a=u,\;s_b=v}}
    \ \sum_{k=a+1}^{b}\Gamma_k(s_{k-1},s_k).
  \label{eq:interval_metric}
\end{equation}
Downward branch-metric accumulation starts from the single-symbol interval metrics, with $G_{k-1:k}(u,v)=\Gamma_k(u,v)$. At each subsequent layer, adjacent interval metrics from the preceding layer are accumulated pairwise. Specifically, for two adjacent intervals $[a,m]$ and $[m,b]$, the accumulated metric over $[a,b]$ is obtained as
\begin{equation}
  G_{a:b}(u,v)=\max_{w\in\mathcal X}
  \left\{G_{a:m}(u,w)+G_{m:b}(w,v)\right\},
  \qquad a<m<b,
  \label{eq:interval_combination}
\end{equation}
where $w$ denotes the intermediate state at time $m$. For each fixed initial state $u$ and final state $v$, the accumulated metrics are evaluated for all possible intermediate states $w$, and the maximum is retained. The accumulations of different interval pairs within the same layer are independent and can therefore be performed in parallel. By repeating this operation layer by layer, the interval length represented by each metric is doubled at each layer. The downward accumulation terminates once the metrics $G_{0:N/2}$ and $G_{N/2:N}$ have been obtained. These two half-block metrics are then used to initialize the forward and backward state metrics at the block midpoint.

In the upward state-metric propagation stage, processing starts from the initialized state metrics at the block midpoint and proceeds to progressively smaller intervals. Together with the block-boundary metrics $A_0$ and $B_N$, the midpoint metrics provide the boundary conditions for the two half-blocks. For an interval $[a,b]$ with available boundary metrics $A_a$ and $B_b$, the state metrics at $m=(a+b)/2$ are calculated as
\begin{equation}
  \begin{aligned}
    A_m(w)
    &=\max_{u,w\in\mathcal X}
      \left\{A_a(u)+G_{a:m}(u,w)\right\},\\
    B_m(w)
    &=\max_{v,w\in\mathcal X}
      \left\{G_{m:b}(w,v)+B_b(v)\right\}.
  \end{aligned}
  \label{eq:midpoint_reconstruction}
\end{equation}
For each midpoint state $w$, the forward update combines the left-boundary metric $A_a(u)$ with the stored interval metric $G_{a:m}(u,w)$. It then selects the maximum over all initial states $u$ to obtain $A_m(w)$. Similarly, the backward update combines $G_{m:b}(w,v)$ with the right-boundary metric $B_b(v)$. It then selects the maximum over all final states $v$ to obtain $B_m(w)$. Both updates use interval metrics obtained during the downward stage. Once $A_m$ and $B_m$ are available, the left subinterval $[a,m]$ receives the boundary metrics $A_a$ and $B_m$, while the right subinterval $[m,b]$ receives the boundary metrics $A_m$ and $B_b$. The midpoint metrics of these subintervals can then be computed independently at the next upward layer. Repeating this procedure halves the interval length at each layer until the forward metric $A$ and backward metric $B$ for all states at the required symbol positions have been obtained. Then, the forward and backward metrics are added according to Eq.~\eqref{eq:state_soft_metric} to obtain the logarithmic symbol posterior.

\begin{table*}[!t] 
  \centering 
  \caption{Comparison of multipliers, adders, comparators, and registers 
  for conventional SMLSE and the proposed layered SMLSE.} 
  \label{tab:smlse_complexity} 
 
  \renewcommand{\arraystretch}{1.5} 
  \setlength{\heavyrulewidth}{1.5pt} 
  \setlength{\lightrulewidth}{0.8pt} 
  \small 

  \begin{tabular*}{10.8cm}{@{\hspace{2mm}\extracolsep{\fill}}ccc@{\hspace{2mm}}}
    \toprule 
    Resource
    & Conventional SMLSE
    & Proposed layered SMLSE \\ 
    \midrule 
 
    Multipliers
    & $16N$
    & $16N$ \\ 
 
    Adders
    & $32N+20D+17$
    & $80N+36D-111$ \\ 
 
    Comparators
    & $12(N+D)$
    & $60N-84$ \\ 
 
    Registers
    & $4(N+D)^2+12N+16D$
    & $32N\log_2N-56N+12D+112$ \\ 
 
    \bottomrule 
  \end{tabular*} 
\end{table*}

\subsection{Processing latency and computational complexity}
\label{sec:smlse_latency}

A sliding-block architecture is adopted for practical implementation \cite{sliding}. Each block consists of $N$ symbols, including $O$ pre-overlap symbols, $D$ useful symbols, and $O$ post-overlap symbols, such that $N=2O+D$. As shown in Fig.~\ref{fig:processing}, the core processing latency is evaluated from the availability of the branch metrics to the completion of the required state-metric calculations. Each recursion step or parallel processing layer is assumed to require one processing cycle. In conventional SMLSE, the forward recursion for $A$ and the backward recursion for $B$ can be performed concurrently. Each recursion traverses its corresponding $O$ overlap symbols and $D$ useful symbols. The total core latency $T_{\mathrm{core}}$ is therefore $O+D$ processing cycles, which scales linearly with $N$. In the proposed layered SMLSE, downward branch-metric accumulation requires $\log_2N-1$ processing cycles to reach the block midpoint, with midpoint initialization incorporated into the final downward processing cycle. Upward state-metric propagation then requires another $\log_2N-1$ processing cycles to obtain both $A$ and $B$ for all states at the central $D$ useful symbol positions. Since the downward and upward stages are performed successively, the total core latency $T_{\mathrm{core}}$ is $2(\log_2N-1)$ processing cycles. The proposed architecture thus reduces the latency from $\mathcal{O}(N)$ to $\mathcal{O}(\log_2 N)$.

The computational complexity is evaluated in terms of multipliers, adders, comparators, and registers. In conventional SMLSE, branch-metric calculation requires $16N$ multipliers and $16N+17$ adders. The generated branch metrics are retained in $4(N+D)^2+8(N+D)$ registers for use in the forward and backward recursions. The calculation of $A$ and $B$ requires $16(N+D)$ adders and $12(N+D)$ comparators, with $4(N+D)$ registers storing the state metrics. Additionally, $4D$ adders and $4D$ registers are used to form and store the symbol soft information for LLR generation. In the proposed layered SMLSE, branch-metric calculation has the same computational complexity as in conventional SMLSE. Downward branch-metric accumulation introduces $64(N-2)$ adders and $48(N-2)$ comparators to combine adjacent interval metrics in parallel. The original branch metrics and composed interval metrics are retained in $32N\log_2N-56N+112$ registers for upward state-metric propagation. The calculation of $A$ and $B$ requires $32D$ adders, $12(N+1)$ comparators, and $8D$ registers. Additionally, $4D$ adders and $4D$ registers are used to form and store the symbol soft information for LLR generation. Table~\ref{tab:smlse_complexity} summarizes the computational complexity. Both architectures require $16N$ multipliers for branch-metric calculation. For the analyzed block length $N$, the proposed architecture uses more adders and comparators but requires fewer registers. Parallel branch-metric accumulation in the downward stage reduces the processing latency at the expense of additional adders and comparators. The reduced processing latency allows the metrics to be buffered for fewer processing cycles, thereby reducing the number of required registers.


\begin{figure}[!t]
\centering
\includegraphics[width=10.8cm]{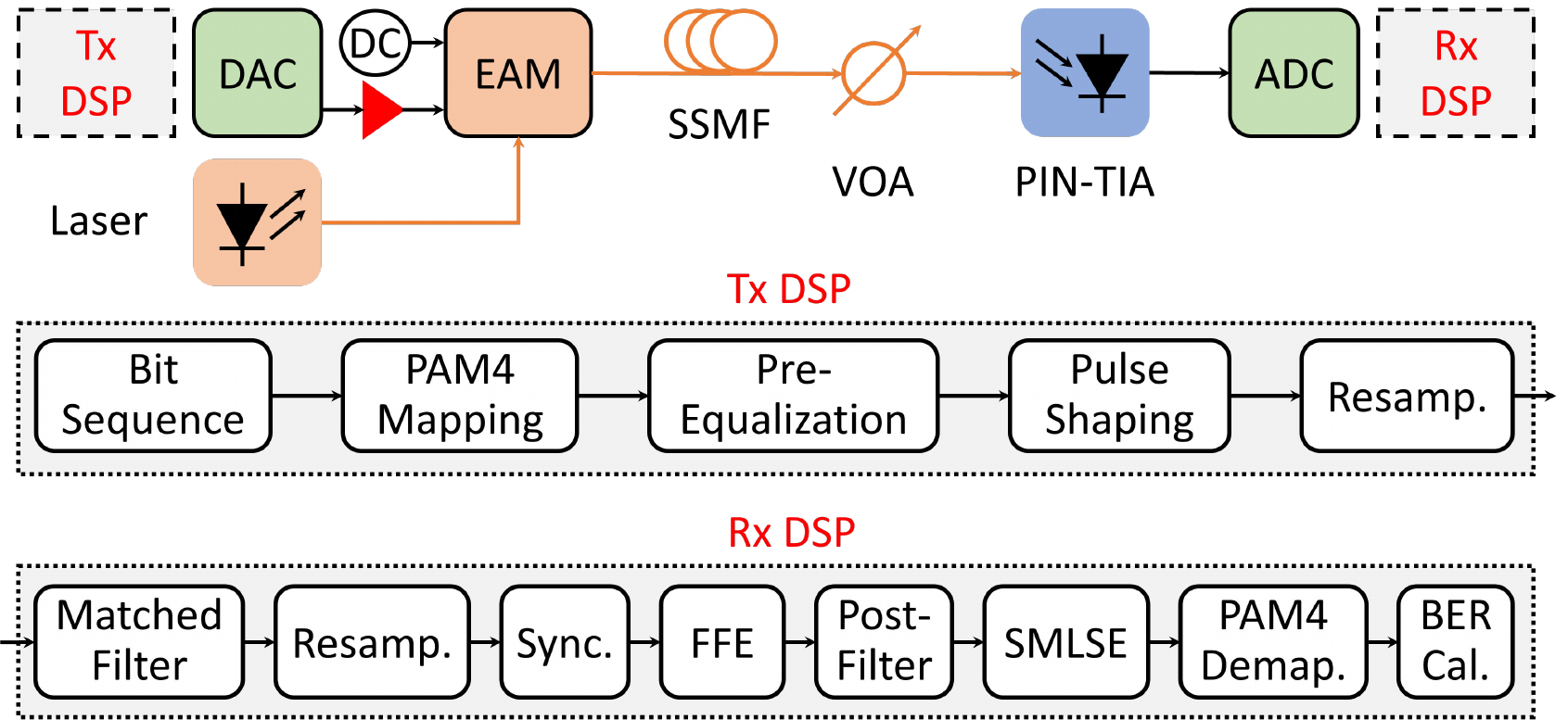}
\caption{Experimental setup and digital signal processing flow for 112-Gbit/s PAM4 transmission over a 2-km SSMF link.}
\label{setup}
\end{figure}

\section{Experimental setup and results}

Figure~\ref{setup} presents the experimental setup for evaluating layered SMLSE through 112-Gbit/s PAM4 transmission over 2 km of SSMF. At the transmitter, $4 \times 10^5$ pseudo-random bits are mapped to $2 \times 10^5$ PAM4 symbols. Optimized pre-equalization \cite{guo} partially compensates for bandwidth-induced distortion, and pulse shaping uses a roll-off factor of 0.1. The waveform is resampled to 1.25 samples per symbol and converted into a 56-GBaud electrical signal by an 8-bit, 70-GSa/s digital-to-analog converter (DAC). A 40-Gbps electrical amplifier (EA, CENTELLAX OA4MVM3) amplifies the DAC output to drive a 40-Gbps electro-absorption modulator (EAM, OM5757C-CTM388). After SSMF transmission, a variable optical attenuator (VOA) adjusts the received optical power (ROP). A 30-Gbps PIN photodiode with an integrated transimpedance amplifier (PIN-TIA) detects the optical signal, and a 70-GSa/s analog-to-digital converter (ADC) digitizes its output. Offline receiver digital signal processing includes matched filtering, resampling, synchronization, FFE, post-filter, the proposed layered SMLSE, PAM4 demapping, and BER calculation.

\begin{figure}[!t]
\centering
\includegraphics[width=12.5cm]{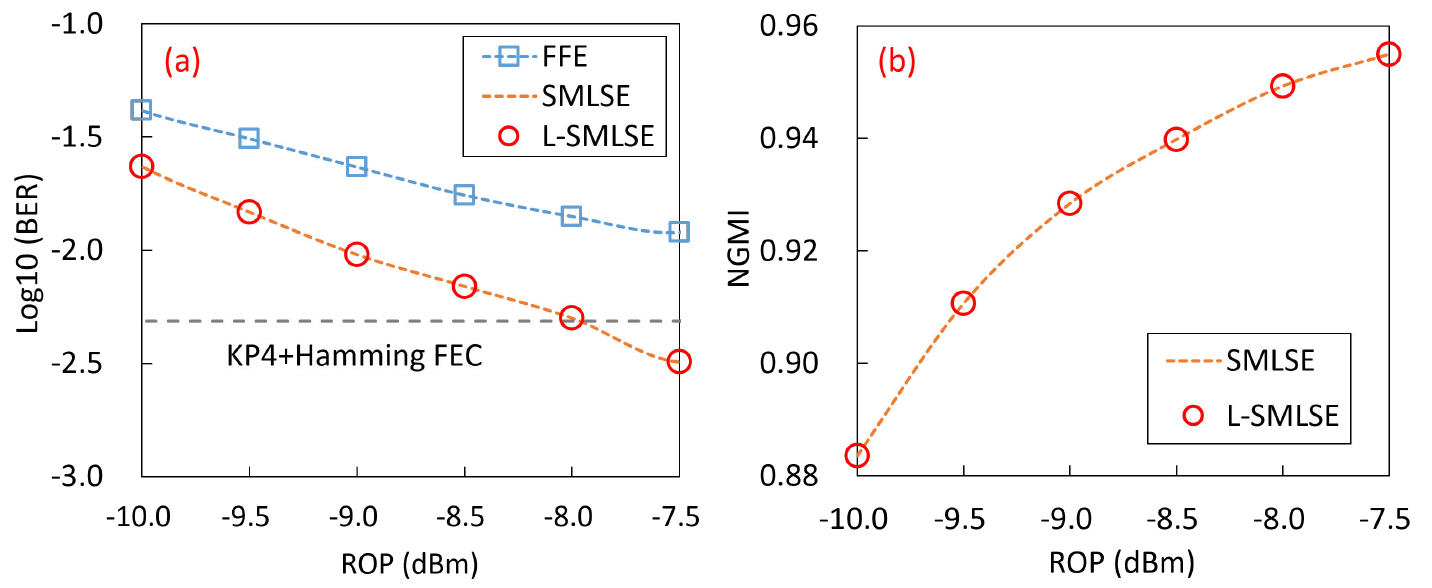}
\caption{(a) BER and (b) NGMI versus ROP for conventional SMLSE and layered SMLSE (L-SMLSE) after 112-Gbit/s PAM4 transmission over 2-km SSMF. The BER performance of FFE is included in (a) for comparison.}
\label{BER}
\end{figure}

Figure~\ref{BER} compares the BER and NGMI performance of conventional SMLSE and layered SMLSE. As shown in Fig.~\ref{BER}(a), both SMLSE schemes provide better BER performance than FFE, and their BER performance is identical. The BER performance of both SMLSE schemes can achieve the KP4+Hamming FEC threshold of $4.85\times10^{-3}$, whereas FFE remains above the threshold throughout the tested ROP range. This soft-output threshold corresponds to the Hamming(128,120)+KP4 concatenated FEC considered for IEEE P802.3dj 800GBASE-FR4, and has also been adopted in recent high-speed IM/DD studies~\cite{Oettinghaus2026DSP400G,Kulmer2026Plasmonic400G}. Fig.~\ref{BER}(b) compares the soft-output performance of conventional SMLSE and layered SMLSE. The NGMI increases from approximately 0.884 to 0.955 as the ROP increases, with no visible difference between the two architectures. These results indicate that layered SMLSE achieves the same BER and NGMI performance as conventional SMLSE.

\begin{figure}[!t]
\centering
\includegraphics[width=12.5cm]{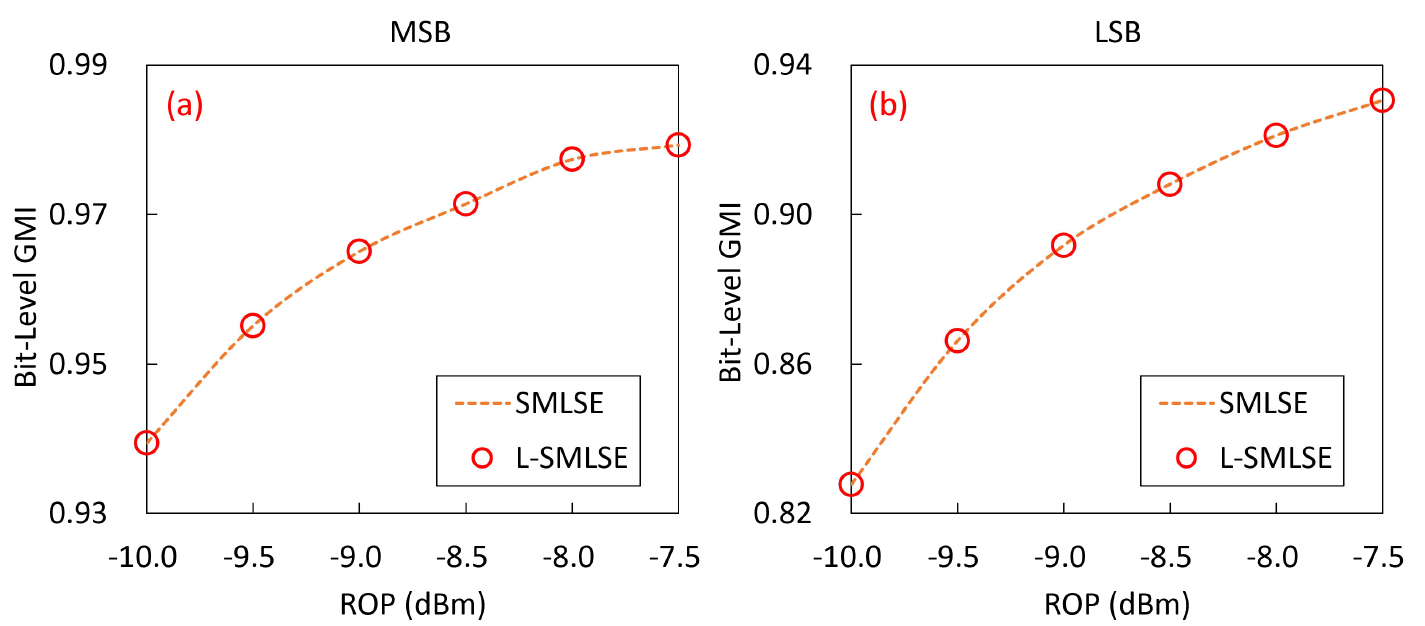}
\caption{Bit-level GMI versus ROP for the (a) MSB and (b) LSB of PAM4 signals using conventional SMLSE and layered SMLSE (L-SMLSE).}
\label{NGMI}
\end{figure}

To further examine the soft-output performance at the individual PAM4 bit level, Fig.~\ref{NGMI} presents the bit-level generalized mutual information (GMI) of the MSB and LSB. The MSB GMI in Fig.~\ref{NGMI}(a) increases from approximately 0.939 to 0.979, whereas the LSB GMI in Fig.~\ref{NGMI}(b) increases from approximately 0.828 to 0.931 over the same ROP range. Thus, the MSB provides more reliable soft information than the LSB. This difference is consistent with the Gray mapping described in Section~\ref{sec:conventional_smlse}, in which the MSB distinguishes positive from negative PAM4 levels, whereas the LSB distinguishes the inner from the outer levels. The adjacent-level errors between $-1$ and $+1$ affect the MSB, while those between $-3$ and $-1$ or between $+1$ and $+3$ affect the LSB. Consequently, the reliability of the MSB is affected near the central decision threshold, whereas that of the LSB is affected near two decision thresholds. For each of the two bit positions, the layered SMLSE results coincide with those of conventional SMLSE. Consequently, the agreement in the overall NGMI is also maintained separately for the two bit positions.

\begin{figure}[!t]
\centering
\includegraphics[width=12.8cm]{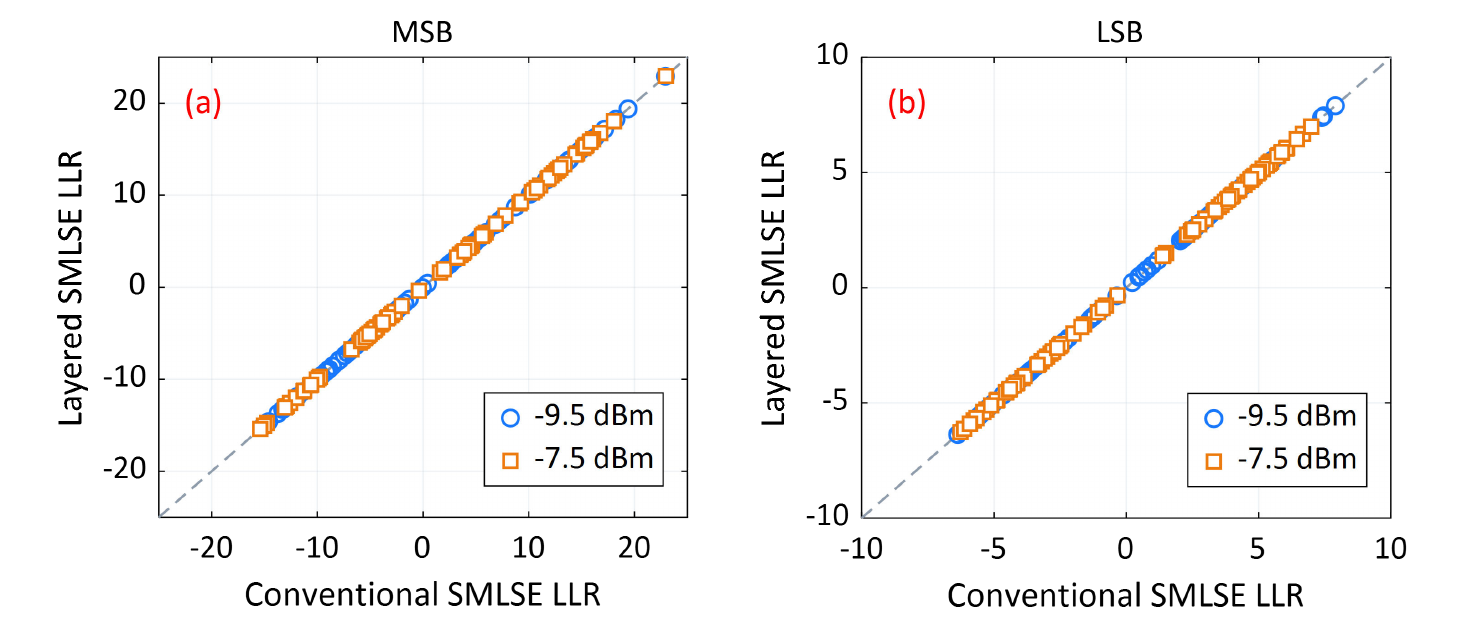}
\caption{Comparison of LLRs obtained by conventional SMLSE and layered SMLSE for the (a) MSB and (b) LSB of PAM4 signals at ROPs of $-9.5$ and $-7.5$~dBm. The dashed line indicates identical LLR values.}
\label{LLR}
\end{figure}

Figure~\ref{LLR} compares individual LLR values from the two architectures at ROPs of $-9.5$ and $-7.5$~dBm, testing the consistency of individual decision reliabilities in addition to the BER and GMI measures. The horizontal and vertical axes represent the conventional and layered SMLSE results, respectively, and the dashed diagonal denotes equal LLR values. For both the MSB in Fig.~\ref{LLR}(a) and the LSB in Fig.~\ref{LLR}(b), the points at both ROPs lie along this diagonal. The experimental results show no difference between the LLR values obtained by layered SMLSE and conventional SMLSE. The LLR sign determines the preferred bit and its magnitude describes decision reliability. The displayed MSB values also span a wider range than the LSB values, consistent with the higher MSB GMI in Fig.~\ref{NGMI}. For the outer PAM4 levels, the nearest symbol with an opposite MSB is farther away than the nearest symbol with an opposite LSB. This increased Euclidean distance can produce larger metric differences, resulting in larger MSB LLR magnitudes. This behavior reflects the bit-dependent distance relationships among the PAM4 symbols. These LLR results indicate that, although the layered architecture changes the order of metric calculations, it produces the same soft outputs as conventional SMLSE. This agreement is consistent with the identical BER and GMI performance observed for the two schemes.

\begin{figure}[!t]
\centering
\includegraphics[width=11cm]{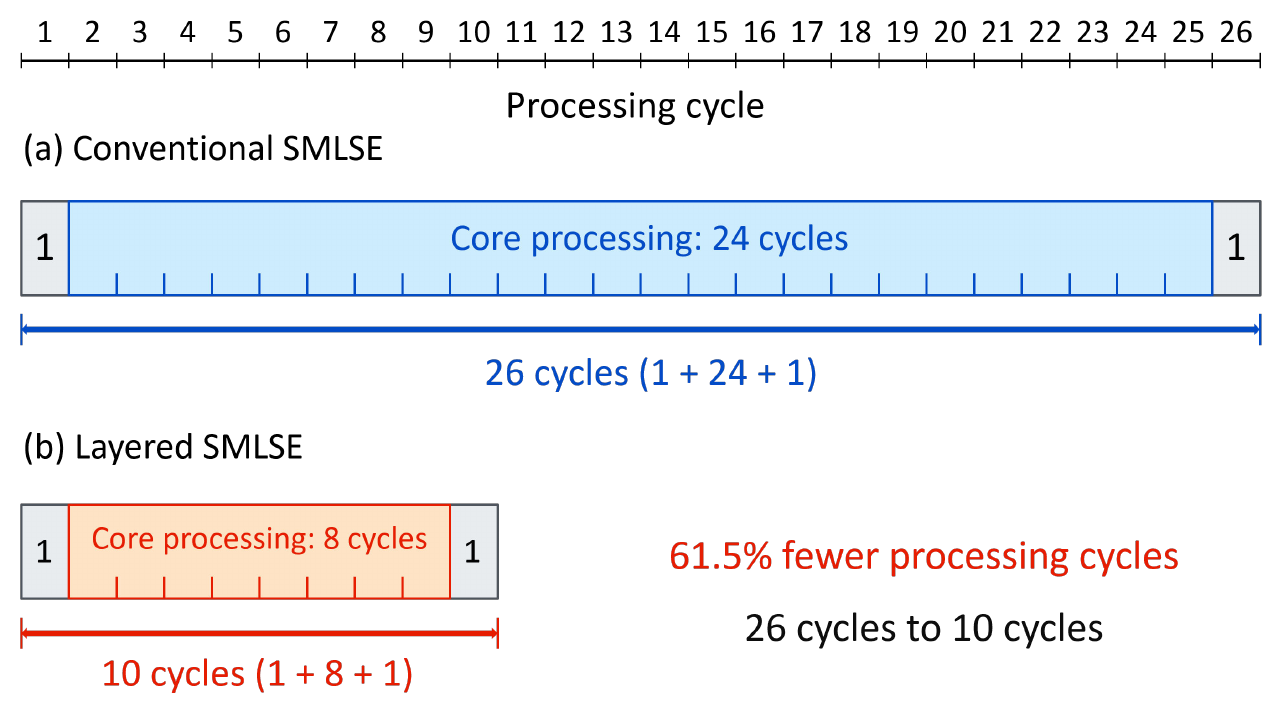}
\caption{Comparison of total processing latency between (a) conventional SMLSE and (b) layered SMLSE, which require 26 and 10 processing cycles, respectively.}
\label{latency}
\end{figure}

The processing latency is compared for a 32-symbol block containing 8 pre-overlap symbols, 16 useful symbols, and 8 post-overlap symbols. Further increasing the number of overlap symbols or useful symbols yields no observable performance improvement under the tested conditions. As shown in Fig.~\ref{latency}(a), under the assumption that each recursion step or parallel layer takes one processing cycle, conventional SMLSE requires 24 cycles for core processing. The forward and backward recursions run concurrently, so their cycle counts are not added. As shown in Fig.~\ref{latency}(b), in layered SMLSE, downward branch-metric accumulation takes 4 cycles, and upward state-metric propagation takes another 4 cycles, yielding a core latency of 8 cycles. With one additional cycle for branch-metric calculation and one for soft-output generation in each architecture, the total latency decreases from 26 to 10 cycles. These results show that layered SMLSE reduces processing latency while retaining the soft-output performance of conventional SMLSE. For the computational complexity, both architectures require 512 multipliers for branch-metric calculations. Layered SMLSE increases the number of adders from 1361 to 3025 and the number of comparators from 576 to 1836 because of the parallel interval-metric combinations. The number of registers decreases from 9856 to 3632 because of the reduced processing latency. In future work, the same simplification techniques can be applied to both architectures to further reduce computational complexity.

\section{Conclusion} \label{conclusion}
In this paper, we propose a low-latency layered SMLSE architecture for PAM4 short-reach optical interconnects. By combining downward branch-metric accumulation with upward state-metric propagation, the architecture reduces the processing latency from linear to logarithmic order while preserving the same soft outputs as conventional SMLSE. Experimental results for 112-Gbit/s PAM4 transmission over 2-km SSMF show identical BER, NGMI, and bit-level LLRs for the two architectures. The total processing latency decreases from 26 to 10 processing cycles. In conclusion, the proposed layered SMLSE architecture enables low-latency soft-output detection for SD-FEC while retaining the decision reliability of conventional SMLSE.

\begin{backmatter}
\bmsection{Funding}
National Key R\&D Program of China (2023YFB2905700); Science Research Project of Hebei Education Department (QN2025007); Hebei Natural Science Foundation (F2026210016); National Natural Science Foundation of China (62371207, 62005102); Young Elite Scientists Sponsorship Program by CAST (2023QNRC001).

\bmsection{Disclosures}
\noindent The authors declare no conflicts of interest.

\end{backmatter}



\bibliography{reference}






\end{document}